\documentclass{emulateapj}
\usepackage{graphicx,amssymb}
\usepackage[usenames,dvips]{color}
\newcommand{\kms}{\ifmmode {\rm km\,s}^{-1} \else km\,s$^{-1}$\fi}

\shorttitle{44~GHz Methanol Masers in the Galactic Center Region}
\shortauthors{Pihlstr\"om et al.}

\begin{document}

\title{Expanded Very Large Array Detection of 44.1~GH\lowercase{z}
  Class I Methanol Masers in Sagittarius A}

\author{Y.~M.~Pihlstr\"om\altaffilmark{1}}
\affil{Department of Physics and Astronomy, University of New Mexico, MSC07 4220, Albuquerque, NM 87131}

\email{ylva@unm.edu}

\altaffiltext{1}{Y.~M.~Pihlstr\"om is also an Adjunct Astronomer at the
  National Radio Astronomy Observatory}

\author{L.~O.~Sjouwerman} \affil{National Radio Astronomy Observatory,
  P.O. Box O, 1003 Lopezville Rd., Socorro, NM 87801}

\and

\author{V. L. Fish}
\affil{MIT Haystack Observatory, Route 40, Westford, MA 01886}

\begin{abstract}
  We report on the detection of 44~GHz Class I methanol (CH$_3$OH)
  maser emission in the Sgr\,A complex with the Expanded Very Large
  Array (EVLA). These EVLA observations show that the Sgr\,A complex
  harbors at least four different tracers of shocked regions in the
  radio regime. The 44~GHz masers correlate with the positions and
  velocities of previously detected 36~GHz CH$_3$OH masers, but less
  with 1720~MHz OH masers. Our detections agree with theoretical
  predictions that the densities and temperatures conducive for
  1720~MHz OH masers may also produce 36 and 44~GHz CH$_3$OH maser
  emission.  However, many 44~GHz masers do not overlap with 36~GHz
  methanol masers, suggesting that 44~GHz masers also arise in regions
  too hot and too dense for 36~GHz masers to form. This agrees with
  the non-detection of 1720~MHz OH masers in the same area, which are
  thought to be excited under even cooler and less dense conditions.
  We speculate that the geometry of the 36~GHz masers outlines the
  current location of a shock front.
\end{abstract}


\section{Introduction}
\label{intro}
The Sagittarius A complex is one of the best studied regions in the
sky and encompasses several interesting phenomena like the nearest
supermassive nuclear black hole (Sgr\,A*), the Circumnuclear Disk (CND), star
forming regions (SFRs) and supernova remnants (SNRs). The line of
sight toward the Sgr\,A complex consists of the SNR Sgr\,A\,East in
the back and the CND (whose ionized part is known as Sgr\,A\,West) in
the front partly overlapping with Sgr\,A\,East. Molecular gas is
abundant and well distributed; the CND consists of irregularly
distributed clumps of molecular gas and there are two giant molecular
cloud cores (GMCs) called the $+$20 and $+50$\,\kms\ clouds
(M$-$0.13$-$0.08 and M$-0.02-0.07$ respectively). These GMCs form the
molecular belt stretching across the Sgr\,A\,complex, providing the
interstellar medium (ISM) that interacts with Sgr\,A\,East. A nice and
recent comprehensive overview is presented by \citet{amo11}.

Bright maser lines are useful probes of physical conditions within
molecular clouds, especially when mapped in detail by
interferometers. One example is the collisionally pumped 1720~MHz OH
maser which is widely recognized as a tracer for shocked regions,
observed both in SFRs and SNRs. In SNRs, Very Large Array (VLA)
observations have shown they originate in regions where the shocks
collide with the interstellar medium
\citep[e.g.][]{claussen97,yusef-zadeh03,frail98}. Such OH masers are
numerous in Sgr\,A\,East, thus probing the conditions of the
interaction regions between the $+50$ and $+$20\,\kms\ clouds and the
SNR Sgr\,A\,East.

Dense gas structures in the Galactic center region, including
Sgr\,A\,East, are traced by ammonia and methanol thermal emission
\citep{coil00,szczepanski89,szczepanski91}. Methanol abundances are
high enough to produce maser emission. Like 1720~MHz OH masers, Class
I methanol masers such as the 36 and 44~GHz transitions are excited
through collisions. Theoretical modeling of collisional OH excitation
predicts that the 1720~MHz OH should be found in regions of $n\geq
10^5$ cm$^{-3}$ and $T\sim 75$\,K
\citep{gray91,gray92,wardle99,lockett99,pihlstrom08}.  The number
density and temperature required for 36~GHz methanol masers are near
those modeled for 1720~MHz~OH masers, with $n\sim 10^4-10^5$ cm$^{-3}$
and $T<100$\,K \citep{morimoto85,cragg92,liechti96}. At least in SFRs,
higher densities and temperatures of $n\sim 10^5-10^6$ cm$^{-3}$ and
$T=80 - 200$\,K will optimize the maser output for the Class I 44~GHz
line, while the 36~GHz maser eventually becomes quenched
\citep{sobolev05,sobolev07,pratap08}.  Finding these methanol maser
lines may therefore help to constrain the upper limit of the density
in the shocked SNR regions. In turn such limits can be used to
estimate the importance of compression by shocks in the formation of
stars near SNRs.

\begin{deluxetable*}{crrcrcrr}
  \tabletypesize{\scriptsize} 
  \tablecaption{44~GHz methanol maser associations}
  \tablewidth{0pt} 
  \tablehead{\colhead{} & \multicolumn{2}{c}{Position} & \colhead{Pointing} &
    \colhead{V$_{LSR}$} & \colhead{T$_{b,44~GHz}$} & \multicolumn{2}{c}{Flux density\tablenotemark{a}} \\
    & \colhead{Right Asc.} & \colhead{Declination} & \colhead{position} & \colhead{\kms}
    & \colhead{10$^3$~K} &
    \multicolumn{2}{c}{Jy beam$^{-1}$} \\
    & \multicolumn{2}{c}{(J2000)} & & & & 44~GHz & 36~GHz \\} \startdata
  1 & 17 45 38.92 & $-$29 00 21.0 & Calibrator & 44.0 & 0.1 & 0.13 &      \\
  2 & 17 45 43.94 & $-$29 00 19.5 & C, Calibrator   & 49.9 & 2.1 & 2.16 &  1.0 \\
  $\phantom{\tablenotemark{b}}$3\tablenotemark{b} & 17 45 49.36 & $-$28 58 53.3 & E & 45.2 & 0.7 & 0.93 & 75.5 \\
  $\phantom{\tablenotemark{c}}$4\tablenotemark{c} & 17 45 49.56 & $-$28 59 00.6 & E & 46.5 & 1.4 & 1.43 & 35.0 \\
  5  & 17 45 49.67 & $-$28 58 55.2 & E & 42.7 & 2.4 & 2.42 &  \\
  6  & 17 45 49.89 & $-$28 59 04.5 & E & 36.9 & 1.4 & 1.47 &  \\
  7  & 17 45 50.48 & $-$29 00 05.0 & D & 49.9 & 0.7 & 0.55 & 36.9 \\
  8  & 17 45 50.48 & $-$28 58 48.7 & E & 59.5 & 0.5 & 0.50 &  \\
  9  & 17 45 50.49 & $-$28 59 07.6 & E & 42.8 & 0.4 & 0.45 &  \\
  10 & 17 45 50.65 & $-$28 59 08.8 & E & 28.0 & 0.7 & 0.74 &  5.2 \\
  11 & 17 45 51.41 & $-$28 59 10.2 & E & 44.2 & 0.5 & 0.45 &  \\
  12 & 17 45 51.55 & $-$28 58 53.7 & E & 46.7 & 0.5 & 0.51 &  \\
  13 & 17 45 51.62 & $-$28 58 26.5 & E & 52.7 & 0.6 & 0.56 &  \\
  14 & 17 45 51.88 & $-$28 58 52.9 & E & 47.1 & 2.3 & 2.39 & 15.9 \\
  $\phantom{\tablenotemark{c}}$15\tablenotemark{c}& 17 45 52.12 & $-$28 58 55.4 & E & 45.9 & 0.6 & 0.64 & 11.3 \\
  16 & 17 45 52.16 & $-$28 58 22.3 & E & 55.4 & 1.1 & 1.11 &  \\
  17 & 17 45 52.43 & $-$28 58 58.2 & E & 43.7 & 0.6 & 0.58 &  \\
  18 & 17 45 52.55 & $-$28 58 56.8 & E & 45.7 & 0.6 & 0.60 &  \\
\enddata
\tablenotetext{a}{Maser peak flux density in the channel of maximum
  flux; highly unreliable for masers 2, 13 and 16 detected beyond
  the primary beam. The 36~GHz peak fluxes are from the data from \citet{sjouwerman10}.}
\tablenotetext{b}{Brightest component in a cluster of detections.}
\tablenotetext{c}{Multiple spectral features detected at the same position.}
\label{masers}
\end{deluxetable*}

That Class I methanol maser lines are detectable in SNR/cloud
interaction regions was shown by \cite{sjouwerman10}, using the 7
first antennas outfitted with 36~GHz receivers at the Expanded VLA
(EVLA, \citet{perley11}).  Several bright masers were found near the
1720~MHz OH masers in the Sgr\,A\,East molecular cloud - SNR
interaction region; a feature also observed by many others
\citep[e.g.][]{tsuboi09}. To test whether the relation between the
collisionally excited 36 and 44 methanol masers and 1720~MHz~OH holds
in general, we here present the result of a search for Class I 44~GHz
methanol maser emission in the Sgr\,A region.

\section{Observations}
On 2010 October 16, the EVLA was used in its C configuration to
observe the $J=7_{0}\rightarrow 6_1 A^{+}$ rotational transition of
CH$_3$OH at 44\,069.41~MHz as part of observing project 10B-146. The
complete results on all sources will be reported elsewhere; here we
concentrate on the emission detected in the Sgr\,A region. The new
Observation Preparation Tool (OPT) was used to schedule the
observations with a bandwidth of 8~MHz in dual polarization. The
bandwidth was split in 256 channels with a resulting velocity
resolution of approximately 0.2 \kms\ over a total velocity coverage
of 51 \kms. At the time of the observations full Doppler tracking was
not available, but instead the sky frequency was calculated at the
beginning of each scheduling block (observing run) and then kept fixed
throughout the observation.  Velocity errors due to uncorrected
Doppler effects over the length of a scan are much less than the
channel width.  We note that both maser and thermal emission may be
detected when observing in the C-configuration (see Sec.\
\ref{results}).

\begin{figure}[th]
\begin{center}
\resizebox{\columnwidth}{!}{\includegraphics{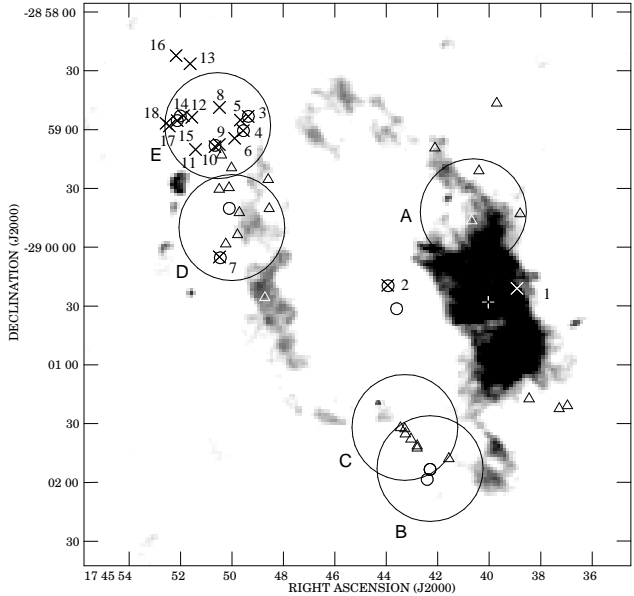}}
\caption{Relative positions of the 44~GHz methanol masers (crosses),
  36~GHz methanol masers (circles) and 1720~MHz OH masers
  (triangles). The big circles show the five main field-of-view
  positions A through E covered. The plus symbol marks the position of Sgr\,A*,
  which is an extra field that could be examined for 44~GHz methanol
  masers.  A blow-up of the upper left region is shown in
  Fig.~\ref{fig2}.  }
\label{fig1}
\end{center}
\end{figure}

The primary beam at 44~GHz is about 56\arcsec, and the full Sgr\,A
region could not be covered by a single pointing. We selected five
pointing positions (``A'' through ``E'', see Fig.~\ref{fig1}) with
central velocities based on previous results on 1720~MHz OH masers and
36~GHz methanol masers. Position A corresponds to a region of
high-velocity 1720~MHz OH masers belonging to the circumnuclear disk,
covering LSR velocities between 106 and 157 \kms. In position B we
previously detected 36~GHz methanol masers at velocities around 23
\kms\ (LSR coverage $-3$ to $+$39 \kms). This pointing position partly
overlaps on the sky with pointing position C which has a central
velocity of 48 \kms\ (LSR coverage 22 -- 74 \kms) based on the
1720~MHz OH masers. Finally, positions D and E correspond to a region
where the 50\,\kms\ molecular cloud interacts with Sgr\,A\,East, and
where 1720\,MHz masers are numerous. The LSR velocity coverage for
these pointing positions was 22 -- 74 \kms. As a bonus we could
include the pointing on our calibrator Sgr\,A* and investigate the
same velocity ranges in this part of the sky.

The data were reduced using standard procedures in AIPS, using 3C\,286
as the flux density calibrator resulting in a typical flux density
uncertainty of $15\%$. In the fields were masers were found, a strong
maser channel was used for self-calibration, with its solutions
applied to all channels. Each cube was CLEANed with robust weighting
down to a level of five times the theoretical rms over a field
approximately twice the primary beam. This was necessary since some
masers appeared in the sidelobes and needed to be accounted for in the
CLEANing process. The resulting typical channel rms noise is 15 -- 20
mJy/beam, and the restoring beam is 1.3\arcsec$\times$0.5\arcsec. Peak fluxes
are corrected for primary beam attenuation using AIPS task PBCOR.

\begin{figure}[th]
\begin{center}
\resizebox{\columnwidth}{!}{\includegraphics{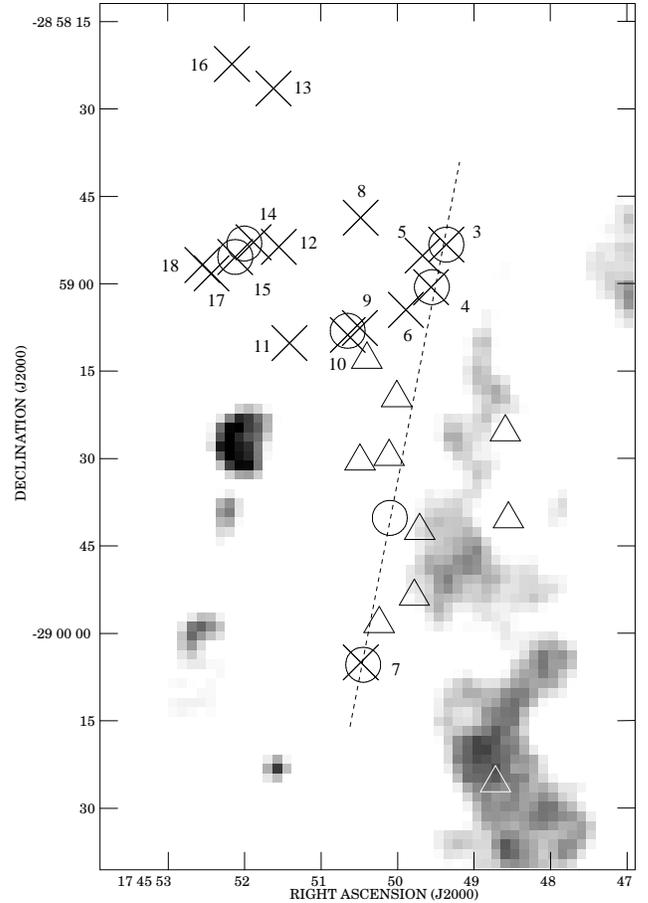}}
\caption{A zoom in of the northeast region of Fig.~\ref{fig1},
  showing an apparent systematic offset in the location of the three
  maser species, with the 44~GHz methanol masers concentrated to the
  northeast, the 1720~MHz OH masers more to the southwest and a
  NNW-SSE line of (four) 36~GHz methanol masers, three co-spatial with
  44~GHz, roughly dividing the two regions. The dashed line shows the
  alignment of 36~GHz masers with the shockfront in the
  NNW-SSE direction.}
\label{fig2}
\end{center}
\end{figure}

\section{Results}
\label{results}
The image cubes were searched for masers, and parameters for the
individual features were extracted using the AIPS task JMFIT. The
combined results are presented in Table \ref{masers} for spectral
features with peak flux densities exceeding 10 times the rms noise. A
few weaker masers exist in the cubes, but they are all located close
to the brighter masers in position and velocity, and will not change
any of the discussion in Sect.\ \ref{discussion}.

A few features show more than one spectral peak at a given position,
implying there is structure on scales smaller than the EVLA beam. Some
spectral features have wings of weaker emission extending over 6 --
8\,\kms. Since the observations were taken when the EVLA was in
C-configuration, we suspect that some of this broad and weak emission
is of thermal origin. The peak flux density of the spectral features
corresponds to brightness temperatures exceeding $10^3$\,K, indicating
that at least some of these features are masers. A definite
confirmation of this needs higher angular resolution
observations. However, several features are co-located spatially and
spectrally with previously detected 36~GHz methanol emission confirmed
to be masers \citep{sjouwerman10}. Both the 44~GHz and 36~GHz masers
are thought to be excited by the same process, so it is likely the
44~GHz observed here is maser emission even though a smaller fraction
of the emission may be of thermal origin.

Three masers were detected outside the primary beam (2, 13 and 16 in
Table \ref{masers}) and their flux density is therefore less certain
than for masers located within the primary beam. The position and
velocity of maser 2 at 50 \kms\ agrees very well with the position of
a 36~GHz maser at 51 \kms\ \citep{sjouwerman10}, and therefore
increases our confidence in this detection. This 44~GHz maser has
previously been reported on by \cite{yusef-zadeh08}, and is associated
with a molecular clump 'G'. Similarly, we confirm their maser 'V' (our
maser 1) in our data taken for the phase-reference calibrator
J1745$-$2900 (i.e.\ Sgr\,A*), and both 'G' and 'V' in archival VLA
data taken on 2009 April 23. The 44~GHz maser associated with clump
'F' reported by \cite{yusef-zadeh08} is not confirmed by our, nor the
archival, observations. The additional two masers detected outside the
primary beam, 13 and 16, also do not have 36~GHz masers directly
associated with them.

Figure \ref{fig1} plots the position of the detected 44~GHz methanol
sources compared to the previously detected 36~GHz and 1720~MHz OH
masers. Also indicated are the spatial regions, primary beams A
through E, selected for maser searches in these observations. Figure
\ref{fig2} shows the northeast region of Sgr\,A\,East where most
44~GHz masers are located. From these plots a few main results can be
concluded. Firstly, the positions and velocities of several 44~GHz
masers (2, 3, 4, 7, 10, 14, and 15 from Table \ref{masers}) agree to
within the errors with the values reported by \cite{sjouwerman10} for
the 36~GHz masers. Secondly, there is a systematic difference between
the overall distributions of the 1720~MHz OH, 36~GHz methanol and
44~GHz methanol masers. In the northeast (upper left) region, in
overlap with the densest part of the 50\,\kms\ cloud, the 44~GHz masers
are offset to the northeast with respect to a narrow, almost linear
southsoutheast to northnorthwest distribution of the 36~GHz
masers. The 1720~MHz OH masers are found on the other side of the
36~GHz masers, near the radio continuum of the SNR to the southwest
(lower right). An offset between the OH and methanol is also observed
in the southeastern interaction region, where the SNR G359.02--0.09
overlaps the Sgr\,A\,East continuum \citep{coil00,herrnstein05}. These
positional offsets are discussed further in Sect.\ \ref{discussion}.

\section{Discussion}
\label{discussion}
To date, four collisionally excited radio frequency maser tracers
have been detected in the Sgr\,A complex; 1720~MHz OH, 36 and 44~GHz
CH$_3$OH and 22~GHz H$_2$O \citep{yusef-zadeh96,
  karlsson03,sjouwerman08,yusef-zadeh08,sjouwerman10}. The 22~GHz
water masers trace regions of higher density and temperature than are
typical for the SNR/cloud interaction regions, and will not be
discussed in this paper. The methanol and hydroxyl are excited under
similar conditions, and in this Letter we discuss their relative
positions and possible origins.

\subsection{36~GHz versus 44~GHz Methanol Masers}
\label{36v44}

Modeling of methanol masers suggest that the 36~GHz transition occurs
under somewhat cooler and less dense ($T\sim$ 30 -- 100\,K, $n\sim
10^4-10^5$ cm$^{-3}$) conditions than the 44~GHz transition ($T\sim$ 80
-- 200 K, $n\sim 10^5-10^6$ cm$^{-3}$; see, e.g.\
\citet{pratap08}). The range of physical conditions do however overlap,
and some spatial overlap could therefore be expected.  Seven 44~GHz
masers (2, 3, 4, 7, 10, 14 and 15) show an almost perfect overlap in
both position and velocity with 36~GHz masers
\citep{sjouwerman10}. Here, according to the modeling, the densities
and temperatures should be close to 10$^5$ cm$^{-3}$ and 100\,K to
produce both methanol maser lines.

It is striking that the brightest 36~GHz masers, all in positions D
and E \citep{sjouwerman10}, are narrowly distributed along a line
roughly from north to south, of which three coincide with 44~GHz
masers (3, 4 and 7) in position and velocity within the errors.  This
NNW-SSE division more or less coincides with the sharp gradient in
low-frequency radio continuum emission of the Sgr\,A\,East SNR
\citep{pedlar89} and appears to be located in the sheath in the CS
emission as mapped by \citet{tsuboi09}.  The mean velocity of each
transition is 46 \kms, implying that they arise in similar regions of
the molecular cloud where the velocities still are less disturbed by
the SNR shock (see Sect.\ \ref{1720comp}). Apart from two individual
exceptions located far from this area (1 and 2), we do not find any
44~GHz (nor 36~GHz) masers westward of this line in our pointings.  It
is therefore tempting to speculate that the line delineates the
arrival of the shock front, where enough material has been swept up to
provide the density for the creation of (very bright 36 GHz) methanol
masers, but not yet enough energy has dissipated to dissociate all
methanol or to significantly disturb the velocity structure by means
of a reverse shock (Section \ref{1720comp}). That the 36~GHz masers
appear to be much brighter than the 44~GHz masers may be due to the
smaller synthesized beams in the 36~GHz observations (200-400 mas
using B-configuration). However, within each observation, the masers
outlining the shock front on average are at least twice as bright as
the other masers in the same transition. It suggests that the geometry
of the shock front, moving in the plane of the sky, causes the path
length amplification to be largest in the line of sight, hence lending
support to our speculation.

In the northeastern part of Sgr\,A\,East toward the core of the
50\,\kms\ cloud, there is a group of 44~GHz masers with a distinct
positional offset from the NNW-SSE line of 36~GHz masers, and which
have no accompanying 36~GHz masers. The narrower distribution of
36~GHz masers suggests that the conditions required to produce masers
in this transition are not fulfilled to the same extent further to the
northeast. The position of the 36~GHz emission is consistent with
being just in the SNR/cloud interaction region, while the 44~GHz
masers may be found deeper inside the denser parts of the cloud where
36~GHz masers are quenched. This situation has been found in sites of
massive star formation, where the 44~GHz masers typically are brighter
than the 36~GHz masers \citep[e.g.][]{pratap08,fish11}, suggesting
some of the 44~GHz masers may be associated with star formation. The
lack of companion 36~GHz masers in this putative star-forming region
in the northeast corner of Sgr\,A\,East \citep[e.g.][]{tsuboi09}
therefore may be due to the limited sensitivity of the 36~GHz
observations.  This picture, at least for this region in the Galactic
center, in which 44~GHz masers are primarily associated with cloud
cores and 36~GHz masers are found at the boundaries of the SNR
interaction regions, is consistent with theoretical models indicating
that 44~GHz masers can be produced at higher densities than 36~GHz
masers.  The existence of 36~GHz masers without accompanying 44~GHz
masers in positions B and C may then indicate the interaction region
of two SNRs without the presence of a dense cloud core.

\subsection{OH versus Methanol Masers}
\label{1720comp}

As is the case with Class I methanol masers, 1720~MHz OH masers are
used as tracers of shocked regions. The presence of 1720~MHz OH masers
indicates the presence of C-shocks \citep[e.g.][]{lockett99}. Modeling
of OH and CH$_3$OH shows that the three maser transitions discussed
here require similar densities and temperatures. This agrees well with
the detection of all three masers in Sgr\,A\,East. However, we observe
a distinct offset in positions between the methanol and OH masers
(Fig.\ \ref{fig1}). In the northeast interaction region between the
50\,\kms\ cloud and Sgr\,A\,East the OH masers are found more to the
southwest. In addition, the 1720~MHz OH masers have a slightly higher
mean velocity of 57~\kms\ versus 46~\kms\ for the
methanol.  We do note however, that the 1720~MHz OH does overlap in
the sky with the line of 36~GHz masers.

A similar offset is observed in the southeastern interaction region
in pointing positions B and C, where the methanol masers are offset
southwest from the OH. The OH mean velocities here are 58 \kms\ to
be compared to the 24.5 \kms\ for the 36~GHz methanol. No 44~GHz
methanol was detected in this region.

The association between 1720~MHz OH and 36~GHz methanol emission may
be due to the processes that form these molecules.  OH is created by
dissociation of H$_2$O (and maybe also CH$_3$OH), and the propagation
of a C-shock creates densities and temperatures suitable for 1720~MHz
OH inversion and probably also some Class I methanol emission
\citep[e.g.][]{lockett99,draine83}. Thus, OH masers should
preferentially be found in the SNR post-shock region. This agrees with
the OH masers being co-located with positions of radio continuum,
outlining regions where electrons have been accelerated by the
shock. The production of methanol is less well understood, but it is
believed that methanol is released from grain mantles, either by
sputtering from a shock or by evaporation when temperatures above
100\,K are reached \citep{hidaka04, menten09, bachiller97, voronkov06,
  hartquist95}. For OH, modeling shows that a temperature around
50-125\,K and a density of 10$^5$cm$^{-3}$ will be conducive to OH
inversion \citep{lockett99,wardle99,pihlstrom08}. This range of values
agrees with both 36~GHz and 44~GHz methanol production, but
non-detections of 6~GHz and higher frequency OH maser transitions in
SNRs like Sgr\,A\,East \citep{fish07,pihlstrom08,mcdonnell08} imply
the densities and temperatures are biased towards the lower values,
perhaps specifically favoring 36~GHz masers to have a closer spatial
connection with 1720~MHz OH masers.

\section{Conclusions}

The methanol and OH masers found in the Sgr\,A complex are located
near each other but are not co-spatial, indicating they trace
different shocks or different regions of the shock. By comparing
velocities of the methanol and the OH we argue that the methanol is
tracing material that has been less disturbed by the shocks, while the
OH is located further in the postshock gas. The narrow, collinear
distribution of 36 GHz methanol in the Sgr\,A\,East interaction region
with the 50\,\kms\ cloud aligns with the SNR shock front, while the OH
is more widely spread into the post-shock regions. 44~GHz and 36~GHz
methanol maser overlaps occur in the cloud/SNR shock region where the
conditions for maser action in both lines are fulfilled. Deeper in the
50\,\kms\ cloud core, there is a group of 44~GHz methanol masers
observed with no accompanying bright 36~GHz masers, indicating a hotter and
denser environment than the material swept up from the shock. These
masers are likely to trace star formation within the cloud core.

The limited number of pointing positions in addition to a limited
velocity coverage makes it hard to in detail determine the
relationship between the methanol and OH masers. New data mapping a
more complete region of Sgr\,A is underway, and we are aiming for a
similar study with a broader velocity range when the EVLA upgrade is
complete.

{\it
  Facilities:} \facility{EVLA}.


\begin{thebibliography}{}

\bibitem[Amo-Baladr{\'o}n et al.(2011)]{amo11} Amo-Baladr{\'o}n, M.~A.,
  Mart{\'{\i}}n-Pintado, J., \& Mart{\'{\i}}n, S.\ 2011, \aap, 526,
  A54

\bibitem[Bachiller \& Perez Gutierrez(1997)]{bachiller97} Bachiller,
  R., \& Perez Gutierrez, M.\ 1997, \apj, 487 L93

\bibitem[Claussen et al.(1997)]{claussen97} Claussen, M.~J., Frail,
  D.~A., Goss, W.~M., \& Gaume, R.~A.\ 1997, \apj, 489, 143

\bibitem[Coil \& Ho(2000)]{coil00} Coil, A.~L., \& Ho, P.~T.~P.\ 2000,
  \apj, 533, 245

\bibitem[Cragg et al.(1992)]{cragg92} Cragg, D.~M., Johns, K.~P.,
  Godfrey, P.~D., \& Brown, R.~D.\ 1992, \mnras, 259, 203

\bibitem[Draine et al.(1983)]{draine83}Draine, B.~T., Roberge, W.~G., \&
  Dalgarno, A.\ 1983, \apj, 264, 485

\bibitem[Fish et al.(2007)]{fish07}Fish, V.~L., Sjouwerman, L.~O., \&
  Pihlstr\"om, Y.~M.\ 2007, \apjl, 670, L117

\bibitem[Fish et al.(2011)]{fish11} Fish, V.~L., Muehlbrad, T.~C.,
  Pratap, P., Sjouwerman, L.~O., Strelnitski, V., Pihlstr\"om, Y.~M., \&
  Bourke, T.~L.\ 2011, \apj, 729, 14

\bibitem[Frail \& Mitchell(1998)]{frail98} Frail, D.A., \& Mitchell,
  G.F.\ 1998, \apj, 508, 690

\bibitem[Gray et al.(1991)]{gray91} Gray, M.~D., Doel, R.~C., \&
  Field, D.\ 1991, \mnras, 262, 30

\bibitem[Gray et al.(1992)]{gray92} Gray, M.~D., Field, D., \&
  Doel, R.~C.\ 1992, \aap, 264, 220

\bibitem[Hartquist et al.(1995)]{hartquist95} Hartquist, T.~W., Menten,
  K.~M., Lepp, S., \& Dalgarno, A.\ 1995, \mnras, 272, 184

\bibitem[Herrnstein \& Ho(2005)]{herrnstein05} Herrnstein, R.~M., \&
  Ho, P.~T.~P.\ 2005, \apj, 620, 287

\bibitem[Hidaka et al.(2004)]{hidaka04} Hidaka, H., Watanabe, N.,
  Shiraki, T.~M. Nagaoka, A., \& Kouchi, A.\ 2004, \apj, 614, 1124

\bibitem[Karlsson et al.(2003)]{karlsson03} Karlsson, R., Sjouwerman,
  L.~O., Sandqvist, A., \& Whiteoak, J.~B.\ 2003, \aap, 403, 1011

\bibitem[Liechti \& Wilson(1996)]{liechti96} Liechti, S., \& Wilson,
  T.L.\ 1996, \aap, 314, 615

\bibitem[Lockett et al.(1999)]{lockett99} Lockett, P., Gauthier, E.,
  \& Elitzur, M.\ 1999, \apjl, 511, L235

\bibitem[McDonnell et al.(2008)]{mcdonnell08}McDonnell, K.~E., Wardle,
  M., \& Vaughan, A.~E.\ 2008, \mnras, 390, 49

\bibitem[Menten et al.(2009)]{menten09} Menten, K.~M., Wilson, R.~W.,
  Leurini, S., \& Schilke, P.\ 2009, \apj, 692, 47

\bibitem[Morimoto et al.(1985)]{morimoto85} Morimoto, M.,
  Kanzawa, T., \& Ohishi, M.\ 1985, \apjl, 288, L11

\bibitem[Pedlar et al.(1989)]{pedlar89} Pedlar, A., Anantharamaiah,
  K.~R., Ekers, R.~D., Goss, W.~M., van Gorkom, J.~H., Schwarz, U.~J.,
  \& Zhao, J.-H.\ 1989, \apj, 342, 769

\bibitem[Perley et al.(2011)]{perley11} Perley, R.~A., Chandler,
  C.~J., Butler, B.~J., \& Wrobel, J.~M.\ 2011, \apjl, in press

\bibitem[Pihlstr\"om et al.(2008)]{pihlstrom08} Pihlstr\"om, Y.~M.,
  Fish, V.~L., Sjouwerman, L.~O., Zschaechner, L.~K., Lockett, P.~B.,
  \& Elitzur, M.\ 2008, \apj, 676, 371

\bibitem[Pratap et al.(2008)]{pratap08} Pratap, P., Shute, P.~A.,
  Keane, T.~C., Battersby, C., \& Sterling, S.\ 2008, \aj, 135, 1718

\bibitem[Szczepanski et al.(1989)]{szczepanski89} Szczepanski, J.~C.,
  Ho, P.~T.~P., Haschick, A.~D., \& Baan, W.~A.\ 1989, IAU Symp.\ 136,
  383

\bibitem[Szczepanski et al.(1991)]{szczepanski91} Szczepanski,
  J.~C., Ho, P.~T.~P., \& Gusten, R.\ 1991, ASP Conf.\ Series, Vol.\ 16,
  143

\bibitem[Sjouwerman et al.(2010)]{sjouwerman10}
  Sjouwerman, L.~O., Pihlstr\"om, Y.~M., \& Fish, V.~L.\ 2010, \apjl, 710,
  L111

\bibitem[Sjouwerman \& Pihlstr\"om(2008)]{sjouwerman08} Sjouwerman,
  L.~O., \& Pihlstr\"om, Y.~M.\ 2008, \apj, 681, 1287

\bibitem[Sobolev et al.(2005)]{sobolev05} Sobolev, A.~M., Ostrovskii,
  A.~B., Kirsanova, M.~S., Shelemei, O.~V., Voronkov, M.~A., \&
  Malyshev, A.~V.\ 2005, in IAU Symp.\ 227, Massive star birth: A
  crossroads of Astrophysics, ed.\ R.\ Cesaroni, M.\ Felli, E.\
  Churchwell \& M.\ Walmsley (Cambridge: Cambridge Univ.\ Press) 174

\bibitem[Sobolev et al.(2007)]{sobolev07} Sobolev, A.~M.\ et al.\
  2007, in IAU Symp.\ 242, Astrophysical Masers and their
  Environments, ed.\ A.\ Vazdekis \& R. Peletier (Cambridge: Cambridge
  Univ.\ Press) 81

\bibitem[Tsuboi et al.(2009)]{tsuboi09} Tsuboi, M., Miyazaki, A., \&
  Okumura, S.~K.\ 2009, \pasj, 61, 29

\bibitem[Voronkov et al.(2006)]{voronkov06} Voronkov, M.~A., Brooks,
  K.~J., Sobolev, A.~M., Ellingsen, S.~P., Ostrovskii, A.~B., \& Caswell,
  J.~L.\ 2006, \mnras, 373, 411

\bibitem[Wardle(1999)]{wardle99} Wardle, M.\ 1999, \apjl, 525, L101

\bibitem[Yusef-Zadeh et al.(2008)]{yusef-zadeh08} Yusef-Zadeh, F.,
  Braatz, J., Wardle, M., \& Roberts, D.\ 2008, \apjl, 683, L147

\bibitem[Yusef-Zadeh et al.(2003)]{yusef-zadeh03} Yusef-Zadeh, F.,
  Wardle, M., Rho, J., \& Sakano, M.\ 2003, \apj, 585, 319

\bibitem[Yusef-Zadeh  et  al.(1996)]{yusef-zadeh96}  Yusef-Zadeh,  F.,
  Roberts, D.~A.,  Goss, W.~M., Frail,  D.~A., \& Green,  A.~J.\ 1996,
  \apjl, 466, L25

\end{thebibliography}
\end{document}